\documentclass[10pt]{article}
\usepackage[square,authoryear]{natbib}
\usepackage{graphicx}
\usepackage{marsden_article}
\usepackage{amsmath,amsthm,amsfonts,amssymb,color,mathrsfs,framed}
\usepackage{epstopdf,framed}
\usepackage{pdfsync}
\usepackage{amscd}
\usepackage{ascmac}

\begin{document}
\newtheorem{remark}[theorem]{Remark}

\title{A free energy Lagrangian variational formulation of the Navier-Stokes-Fourier system}
\vspace{-0.2in}

\newcommand{\todoFGB}[1]{\vspace{5 mm}\par \noindent
\framebox{\begin{minipage}[c]{0.95 \textwidth} \color{red}FGB: \tt #1
\end{minipage}}\vspace{5 mm}\par}


\author{Fran\c{c}ois Gay-Balmaz$^{1}$ and Hiroaki Yoshimura$^{2}$}
\addtocounter{footnote}{1} 
\footnotetext{\'Ecole Normale Sup\'erieure/CNRS, Laboratoire de M\'et\'eorologie Dynamique, 24 Rue Lhomond 75005 Paris, France. \texttt{gaybalma@lmd.ens.fr}}
\addtocounter{footnote}{1}
\footnotetext{School of Science and Engineering, Waseda University Dynamique, Okubo, Shinjuku, Tokyo 169-8555, Japan. 
\texttt{yoshimura@waseda.jp}}

\date{}


\maketitle

\begin{center}
\abstract{We present a variational formulation for the Navier-Stokes-Fourier system based on a free energy Lagrangian. This formulation is a systematic infinite dimensional extension of the variational approach to the thermodynamics of discrete systems using the free energy, which complements the Lagrangian variational formulation using the internal energy developed in \cite{GBYo2016b} as one employs temperature, rather than entropy, as an independent variable. The variational derivation is first expressed in the material (or Lagrangian) representation, from which the spatial (or Eulerian) representation is deduced. The variational framework is intrinsically written in a differential-geometric form that allows the treatment of the Navier-Stokes-Fourier system on Riemannian manifolds.}

\end{center}

\section{Introduction}

The dynamics of a  viscous heat conducting fluid is governed by the Navier-Stokes-Fourier equations given by a system of PDEs describing the balance of fluid momentum, the balance of mass, and the conservation of energy. The latter can be equivalently formulated in terms of the entropy or the temperature. It is well-known that in absence of the irreversible processes of viscosity and heat conduction, these equations, as well as the general equations of reversible continuum mechanics, arise from the Hamilton principle applied to the Lagrangian trajectory of fluid particles.

In \cite{GBYo2016a,GBYo2016b}, we proposed a systematic extension of Hamilton's principle to include irreversible processes by introducing the concept of thermodynamic displacements and making use of a generalization of the Lagrange-d'Alembert principle with nonlinear nonholonomic constraints. This approach covers both discrete and continuum systems and naturally involves the entropy as an independent variable. 
For a concrete use in applications, it is often more practical to use the temperature  rather than the entropy as the independent variable. Temperature is indeed a much easier measurable quantity and the phenomenological coefficients (such as heat conductivity or viscosity) are naturally expressed in terms of the temperature rather than the entropy. In this case, the variational formulation must be expressed in terms of the free energy.

In this paper, we present a variational formulation for the Navier-Stokes-Fourier system based on a {\it free energy Lagrangian}, which complements the approach developed in \cite{GBYo2016b}. The variational derivation is first expressed in the material (or Lagrangian) description, from which the spatial (or Eulerian) description is deduced.  The variational formulation follows from an infinite dimensional extension of the free energy Lagrangian variational formulation for nonequilibrium  thermodynamics of discrete systems. It has a systematic structure which relies on the concepts of variational and phenomenological constraints.

\section{Discrete systems and a free energy Lagrangian}\label{section_2}

In this section we review the variational formulation for nonequilibrium thermodynamics of discrete (i.e., finite dimensional) systems developed in \cite{GBYo2016a}. The formulation is first given in terms of ``classical Lagrangians", i.e., Lagrangians expressed in terms of the internal energy of the system. This naturally implies the use of the entropy $S$ as an independent variable in the variational formulation. Then, we present a variational formulation based on a \textit{free energy Lagrangian} that allows the treatment of the temperature $T$ rather than the entropy as the independent variable.

\subsection{Variational formulation of nonequilibrium thermodynamics of simple systems}

We shall present the variational formulation by first considering \textit{simple thermodynamic systems} before going into the general setting of the discrete systems. We follow the systematic treatment of thermodynamic systems presented in \cite{StSc1974}, to which we also refer for the precise statement of the two laws of thermodynamics.

\medskip

\noindent \textbf{Simple discrete systems.}
A \textit{discrete thermodynamic system} $ \boldsymbol{\Sigma} $ is a collection $ \boldsymbol{\Sigma} = \cup_{A=1}^N$ of a finite number of interacting simple thermodynamic systems $ \boldsymbol{\Sigma} _A $. By definition, a \textit{simple thermodynamic system} is a macroscopic system for which one (scalar) thermal variable and a finite set of mechanical variables are sufficient to describe entirely the state of the system. From the second law of thermodynamics (e.g., \cite{StSc1974}), we can always choose  the entropy $S$ as a thermal variable. A typical example of such a simple system is the one-cylinder problem. We refer to \cite{Gr1999} for a systematic treatment of this system via Stueckelberg's approach. 

\medskip

\noindent \textbf{Variational formulation.} We now quickly review from \cite{GBYo2016a} the variational formulation of nonequilibrium thermodynamics for the particular case of simple closed systems.

Let $Q$ be the configuration manifold associated to the mechanical variables of the simple system. We denote by $TQ$ the tangent bundle to $Q$ and use the classical local notation $(q, v) \in TQ$ for the elements in the tangent bundle. Our approach is of course completely intrinsic and does not depend on the choice of coordinates on $Q$.
The Lagrangian of a simple thermodynamic system is a function
\[
L: TQ \times \mathbb{R}  \rightarrow \mathbb{R} , \quad (q, v, S) \mapsto L(q, v, S),
\]
where $S \in\mathbb{R}$ is the entropy. We assume that the system involves exterior and friction forces given by fiber preserving maps $F^{\rm ext}, F^{\rm fr}:TQ\times \mathbb{R} \rightarrow T^* Q$, i.e., such that $F^{\rm fr}(q, v, S)\in T^*_qQ$, similarly for $F^{\rm ext}$, where $T^*Q$ is the cotangent bundle to $Q$. Finally we assume that the system is subject to an external heat power supply $P^{\rm ext}_H(t)$. 

\medskip

We say that a curve $(q(t),S(t)) \in Q \times \mathbb{R}$, $t \in [t _1 , t _2 ] \subset \mathbb{R}$ is a {\it solution of the variational formulation of nonequilibrium thermodynamics} if it satisfies the variational condition 
\begin{equation}\label{LdA_thermo_simple} 
\delta \int_{t _1 }^{ t _2}L(q , \dot q , S)dt +\int_{t_1}^{t_2}\left\langle F^{\rm ext}(q, \dot q, S), \delta q\right\rangle dt =0, \quad\;\;\; \textsc{\small Variational Condition}
\end{equation}
for all variations $ \delta q(t) $ and $\delta S(t)$ subject to the constraint
\begin{equation}\label{CV_simple} 
\frac{\partial L}{\partial S}(q, \dot q, S)\delta S= \left\langle F^{\rm fr}(q , \dot q , S),\delta q \right\rangle,\qquad\qquad\, \textsc{\small Variational Constraint}
\end{equation}
with $ \delta q(t_1)=\delta (t_2)=0$, and also if it satisfies the phenomenological  constraint 
\begin{equation}\label{CK_simple} 
\frac{\partial L}{\partial S}(q, \dot q, S)\dot S  = \left\langle F^{\rm fr}(q, \dot q, S) , \dot q \right\rangle- P^{\rm ext}_H, \quad \textsc{\small Phenomenological Constraint}
\end{equation}
where $ \dot q=\frac{dq}{dt}$ and $\dot S=\frac{dS}{dt}$.

From this variational formulation, we deduce the system of evolution equations for the simple thermodynamic system as
\begin{equation}\label{simple_systems} 
\left\{
\begin{array}{l}
\displaystyle\vspace{0.2cm}\frac{d}{dt}\frac{\partial L}{\partial \dot q}- \frac{\partial L}{\partial q}=  F^{\rm fr}(q, \dot q, S)+F^{\rm ext}(q, \dot q, S),\\
\displaystyle\frac{\partial L}{\partial S}\dot S= \left\langle F^{\rm fr}(q, \dot q, S), \dot q \right\rangle - P^{\rm ext}_H.
\end{array} \right.
\end{equation} 

\begin{remark}[Phenomenological and variational constraints]\label{PC_VS_VC}{\rm 
The explicit expression of the constraint \eqref{CK_simple} involves phenomenological laws for the friction
force $F^{\rm fr}$, this is why we refer to it as a \textit{phenomenological constraint}. The associated constraint \eqref{CV_simple}  is called a \textit{variational
constraint} since it is a condition on the variations to be used in \eqref{LdA_thermo_simple}. Note that the constraint \eqref{CK_simple} is nonlinear and also that one passes from the variational constraint to the phenomenological constraint by formally replacing the variations $ \delta q$, $\delta S$ by the time derivatives $ \dot q$, $\dot S$. Such a systematic correspondence between the phenomenological and variational constraints still holds for the general discrete systems, as we shall recall below. We refer to \cite{GBYo2017a} for the relation with other variational formulations used in nonholonomic mechanics with linear or nonlinear constraints. See also \S\ref{3_2} below.
}\end{remark}

For the case of adiabatically closed systems (i.e., $P^{\rm ext}_H=0$), the evolution equations \eqref{simple_systems} can be geometrically formulated in terms of Dirac structures induced from the phenomenological constraint and from the canonical symplectic form on $T ^\ast Q$ or on $T ^\ast (Q\times \mathbb{R}  )$, see \cite{GBYo2017c}.

In absence of the entropy variable $S$, this variational formulation recovers Hamilton's variational principle in classical mechanics, where \eqref{simple_systems} becomes the Euler-Lagrange equations.

\subsection{Variational formulation of nonequilibrium thermodynamics of discrete systems}\label{2_2}

\noindent \textbf{Discrete systems.} We now consider the case of a discrete system $ \boldsymbol{\Sigma}  = \cup_{A=1}^N \boldsymbol{\Sigma}  _A$, composed of interconnecting simple systems $ \boldsymbol{\Sigma}  _A$, $A=1,...,N$ that can exchange heat and mechanical power, and interact with \textit{external heat sources} $ \boldsymbol{\Sigma}  _R$, $R=1,...,M$. We follow the description of discrete systems given in \cite{StSc1974} and \cite{Gr1997}.

By definition, a {\it heat source} is a simple system $ \boldsymbol{\Sigma} _R$ uniquely defined by a single variable $S_R$. Its energy is thus given by $U_R=U_R(S_R)$, the temperature is $T^R:= \frac{\partial U_R}{\partial S_R}$, and $ \frac{d}{dt} U_R=T^R\dot S_R=P^{R \rightarrow \boldsymbol{\Sigma}  }_H$, where $P_H^{R \rightarrow \boldsymbol{\Sigma}  }$ is the heat power flow due to the heat exchange with $ \boldsymbol{\Sigma}  $.

The state of the discrete system $ \boldsymbol{\Sigma} $ is described by geometric variables $q \in Q_{ \boldsymbol{\Sigma} } $ and entropy variables $S_A$, $A=1,...,N$. Note that the entropy $S_A$ has the index $A$ since it is associated to the simple system $ \boldsymbol{\Sigma}  _A$. The geometric variables, however, are not indexed by $A$ since in general they are associated to several systems $ \boldsymbol{\Sigma}  _A$ that can interact with.
The Lagrangian of a discrete system is thus a function
\begin{equation}\label{Lagrangian_S_discrete} 
L:T Q_{ \boldsymbol{\Sigma} } \times \mathbb{R}^N \rightarrow \mathbb{R}  , \quad (q, \dot q, S_1,...,S_N) \mapsto L(q, \dot q, S_1,...,S_N).
\end{equation} 

As before, the power supplied from the exterior is due to that by external forces and by transfer of heat. For simplicity, we ignore internal and external matter exchanges in this section. Hence, in particular, we consider the case in which the system is closed.
The external force reads $F^{\rm ext}:=\sum_{A=1}^N F^{{\rm ext} \rightarrow A}$, where $F^{{\rm ext} \rightarrow A}$ is the external force acting on the system $\boldsymbol{\Sigma}  _A$. The external heat power associated to heat transfer is $P^{\rm ext}_H =\sum_R \big(\sum_{A=1}^NP_H^{R \rightarrow A}\big)=\sum_{A=1}^NP_H^{{\rm ext} \rightarrow A}$,
where $P_H^{R \rightarrow A}$ denotes the power of heat transfer  between the external heat source $\boldsymbol{\Sigma}  _R$ and the system $ \boldsymbol{\Sigma}  _A$.
The friction force associated to system $\boldsymbol{\Sigma} _A$ is  $F^{{\rm fr}(A)}:TQ_{ \boldsymbol{\Sigma}  } \times \mathbb{R}^N \rightarrow T^* Q_ {\boldsymbol{\Sigma} }$ with $F^{\rm fr}:=\sum_{A=1}^N F^{{\rm fr}(A)}$. The internal heat power exchange between $ {\boldsymbol{\Sigma} } _A$ and $ {\boldsymbol{\Sigma}} _B$ can be described by
\[
P_H^{B \rightarrow A}= \kappa _{AB}(q , S ^A , S ^B)(T^B-T^A),
\]
where $ \kappa _{AB}= \kappa _{BA}\geq 0$ are the heat transfer phenomenological coefficients. 

A typical, and historically relevant, example of a discrete (non-simple) system is the {\it adiabatic piston}. We refer to \cite{Gr1999} for a systematic treatment of the adiabatic piston from Stueckelberg's approach.

\medskip

\noindent \textbf{Variational formulation.}
Our variational formulation is based on the introduction of new variables, called \textit{thermodynamic displacements}, that allow a systematic inclusion of all the irreversible processes involved in the system. In our case, since we only consider the irreversible processes of the mechanical friction and heat conduction, we just need to introduce (in addition to the mechanical displacement $q$) the \textit{thermal displacements}$^{1}$, $ \Gamma ^A $, $A=1,...,N$ such that  $\dot{\Gamma}^{A}=T^{A}$, where $ \Gamma ^A $ are monotonically increasing real functions of time $t$ and  hence the temperatures $T^{A}$ of $ \boldsymbol{\Sigma} _A$ take positive real values, i.e., $(T^{1},...,T^{N}) \in \mathbb{R}  _{+}^{N}$. \addtocounter{footnote}{1}
\footnotetext{The notion of thermal displacement was first used by \cite{He1884} and in the continuum setting by \cite{GrNa1991}. We refer to the Appendix of \cite{Po2009} for an historical account.}Each of these variables is accompanied with its dual variable $ \Sigma _A$ whose time rate of change is associated to the entropy production of the simple system $A$. The meaning of the variable $\Sigma_A$ and its distinction with the entropy variable $S_A$ may be clarified in the context of continuum systems, as will be seen in \S\ref{3}.

\medskip

We say that a curve $\left(q(t), S _A (t), \Gamma ^A (t) , \Sigma _A (t) \right)  \in Q _{ \boldsymbol{\Sigma} }\times \mathbb{R}^{3N} $, $t \in [t _1 , t _2 ] \subset \mathbb{R}$ is \textit{solution of the variational formulation of nonequilibrium thermodynamics} if it satisfies the \textit{variational condition}
\begin{equation}\label{LdA_thermo_discrete_systems} 
\delta \int_{ t _1 }^{ t _2 }\Big[ L(q  , \dot q , S _1 , ... S _N )+ \sum_{A=1}^N(S_A - \Sigma_A)\dot{\Gamma}^A \Big]   dt +\int_{ t _1 }^{ t _2 } \left\langle F^{\rm ext}, \delta q \right\rangle dt =0,
\end{equation}
for all variations $\delta q (t), \delta \Gamma^A (t)  , \delta \Sigma_A (t)$ subject to the \textit{variational constraint}
\begin{equation}\label{Virtual_Constraints_Systems} 
\begin{split}
\frac{\partial L}{\partial S_A } \delta \Sigma _A
=  \left\langle F^{{\rm fr}(A)}, \delta q  \right\rangle-\sum_{B=1}^N \kappa _{AB}&(\delta \Gamma  ^B-\delta \Gamma ^A ), \;\;\textrm{$($no sum on $A)$}
\end{split}
\end{equation}
with $\delta q( t _i )=0$ and $ \delta \Gamma ( t _i )=0$, for $i=1,2$, and also if it satisfies the nonlinear \textit{phenomenological constraint}
\begin{equation}\label{Kinematic_Constraints_Systems} 
\begin{split}
\frac{\partial L}{\partial S _A }  \dot{\Sigma}_A= \left\langle F^{{\rm fr}(A)}, \dot q  \right\rangle - \sum_{B=1}^N \kappa _{AB}&(\dot \Gamma  ^B-\dot \Gamma ^A ) -  P_H^{\rm ext}. \\ 
\end{split}
\end{equation} 
From this variational formulation, we deduce the system of evolution equations for the discrete thermodynamic system as
\begin{equation}\label{discrete_systems} 
\left\{
\begin{array}{l}
\displaystyle\vspace{0.2cm}\frac{d}{dt}\frac{\partial L}{\partial \dot q}- \frac{\partial L}{\partial q}= \sum_{A=1}^N  F^{{\rm fr}(A)}+ F^{\rm ext},\\
\displaystyle\frac{\partial L}{\partial S_A}\dot S_A= \left\langle F^{{\rm fr}(A)}, \dot q \right\rangle +\sum_{B=1}^N \kappa _{AB} \left(  \frac{\partial L}{\partial S_B}-\frac{\partial L}{\partial S_A}\right) - P^{\rm ext}_H, \;\; A=1,...,N.
\end{array} \right.
\end{equation} 
We refer to \cite{GBYo2016a} for a complete treatment of discrete systems. In a similar way with the situation of simple thermodynamic systems, one passes from the variational constraint \eqref{Virtual_Constraints_Systems} to the phenomenological constraint \eqref{Kinematic_Constraints_Systems} by formally replacing the $ \delta $-variations $ \delta q, \delta \Sigma _A,\delta \Gamma _A$ by the time derivatives $\dot q, \dot \Sigma _A, \dot\Gamma _A$ (see Remark \ref{PC_VS_VC}). This is possible thanks to the introduction of the thermodynamic displacements $\Gamma _A$.

\subsection{Formulation based on the free energy and heat equations}\label{2_3} 

We now present the variational formulation for discrete thermodynamic systems based on a free energy Lagrangian, in which one makes use of temperature $T$ rather than entropy $S$ as an independent variable. We start with the case of a simple discrete thermodynamic system.

\medskip

\noindent \textbf{Simple systems.}
Given the Lagrangian $L:TQ \times \mathbb{R} \rightarrow \mathbb{R}  $ of a discrete simple system, the associated \textit{free energy Lagrangian} $ \mathcal{L} :TQ \times \mathbb{R}_+  \rightarrow \mathbb{R}$ is defined (see \cite{GBYo2016a}) by
\[
\mathcal{L}(q,\dot q,T):= L(q,\dot q, S(q, \dot q, T))+TS(q, \dot q, T),
\]
where we assumed that the function $S  \in \mathbb{R}  \mapsto \frac{\partial L}{\partial S}(q,\dot q, S) \in \mathbb{R}  _+ $ is a diffeomorphism for all $(q, \dot q) \in TQ$ and where the function $S(q,\dot q,T)$ is defined by the condition $ -\frac{\partial L}{\partial S}(q,\dot q, S)=T$, for all $(q,\dot q, S) \in TQ \times \mathbb{R}$.
\addtocounter{footnote}{1}\footnotetext{More strictly speaking, temperature is defined by $T= -\frac{\partial L}{\partial S} \in T_{\Gamma}\mathbb{R} \cong \mathbb{R}_{+}$ for $S \in T^{\ast}\mathbb{R} \cong \mathbb{R}^{\ast}$, or conversely, entropy by $S= \frac{\partial \mathcal{L}}{\partial T} \in T_{\Gamma}^{\ast}\mathbb{R} \cong \mathbb{R}^{\ast}$ for $T \in T_{\Gamma}\mathbb{R} \cong \mathbb{R}_{+}$.}

In most physical examples, the partial derivative $ \frac{\partial L}{\partial S}$ does not depend on $\dot q$. In this case the Lagrangian has the standard form $L(q, \dot q, S)=L_{\rm mech}(q, \dot q)- U(q,S)$, where $L_{\rm mech}: TQ \rightarrow \mathbb{R}  $ is a Lagrangian of the mechanical part of the simple system and $U:Q \times \mathbb{R} \rightarrow \mathbb{R}$ is an internal energy. The associated free energy Lagrangian is $ \mathcal{L} (q, \dot q, T)= L_{\rm mech}(q, \dot q)-  \mathscr{F}(q,T)$, where $\mathscr{F}:Q\times \mathbb{R} _+\rightarrow \mathbb{R} $ is the free energy associated to $U:Q\times\mathbb{R} \rightarrow \mathbb{R} $, where the relation between temperature and entropy may be understood in the dual context of Lagrangians$^{2}$.

The friction and external forces $F^{\rm fr}, F^{\rm ext}: T Q \times \mathbb{R} _+\rightarrow T^*Q$ are now expressed in terms of the temperature rather than the entropy. As before, the variational formulation needs the introduction of the variables $ \Gamma $ and $ \Sigma $.
\medskip

Recall that $T=\dot{\Gamma}$. We say that a curve $(q(t), \Gamma (t), \Sigma (t)) \in Q \times \mathbb{R}^{2}$, $t \in [t _1 , t _2 ] \subset \mathbb{R}$ is a {\it solution of the free energy variational formulation of nonequilibrium thermodynamics} if it satisfies the \textit{variational condition}
\begin{equation}\label{LdA_thermo_discrete} 
\delta \int_{t _1 }^{ t _2} \left( \mathcal{L} (q , \dot q , \dot\Gamma )- \Sigma \dot\Gamma \right) dt +\int_{t_1}^{t_2}\left\langle F^{\rm ext}(q, \dot q, T), \delta q\right\rangle dt =0,
\end{equation}
for all variations $ \delta q(t), \delta \Gamma(t)$ and $\delta \Sigma(t)$ subject to the \textit{variational constraint}
\begin{equation}\label{CV_discrete} 
\dot\Gamma \delta \Sigma = - \left\langle F^{\rm fr}(q , \dot q , T),\delta q \right\rangle,
\end{equation}
with $ \delta q(t_1)=\delta\Gamma (t_2)=0$, and also if it satisfies the nonlinear nonholonomic \textit{phenomenological constraint}
\begin{equation}\label{CK_discrete} 
\dot\Gamma \dot \Sigma  =  - \left\langle F^{\rm fr}(q, \dot q, T) , \dot q \right\rangle + P^{\rm ext}_H.
\end{equation}
From this variational formulation, we get the system of evolution equations for the simple system in terms of the free energy as
\begin{equation}\label{discrete_systems_free} 
\left\{
\begin{array}{l}
\displaystyle\vspace{0.2cm}\frac{d}{dt}\frac{\partial  \mathcal{L} }{\partial \dot q}- \frac{\partial  \mathcal{L} }{\partial q}=  F^{\rm fr}(q, \dot q, T)+F^{\rm ext}(q, \dot q, T),\\
\displaystyle T \frac{d}{dt} \frac{\partial \mathcal{L} }{\partial T}= - \left\langle F^{\rm fr}(q, \dot q, T), \dot q \right\rangle + P^{\rm ext}_H.
\end{array} \right.
\end{equation} 
This is an appropriate form to derive the heat equation of the simple system. Since the Lagrangian has the form $ \mathcal{L} (q, \dot q, T)= L_{\rm mech}(q, \dot q)- \mathscr{F}(q,T)$, the second equation in \eqref{discrete_systems_free} becomes
\[
T \left( \frac{\partial ^2 \mathscr{F}}{\partial T ^2 }(q,T)\dot T+ \frac{\partial  ^2 \mathscr{F}}{\partial q \partial T}(q,T)\dot q \right) =\left\langle F^{\rm fr}(q, \dot q, T), \dot q \right\rangle - P^{\rm ext}_H, 
\]
from which we obtain the \textit{heat equation}
\[
c_v(q,T) \dot T=T \frac{\partial  ^2 \mathscr{F}}{\partial q \partial T}(q,T)\dot q - \left\langle F^{\rm fr}(q, \dot q, T), \dot q \right\rangle+ P^{\rm ext}_H ,
\]
where $c_v(q,T)= - T \frac{\partial ^2\mathscr{F}}{\partial T ^2 }(q,T)$ is the specific heat.

\medskip

\noindent \textbf{Discrete systems.} Consider a discrete system with configuration space $Q_{ \boldsymbol{\Sigma} }$ and Lagrangian $L: T Q_{ \boldsymbol{\Sigma} } \times \mathbb{R}   ^N \rightarrow \mathbb{R}  $. The corresponding free energy Lagrangian is defined by generalizing the above definition as
\begin{align*} 
\mathcal{L} (q, \dot q, T^1,...,T^N):&= L \left( q, \dot q, S_1(q, \dot q, T^1,...,T^N),..., S_N(q, \dot q, T^1,...,T^N)\right) \\
&\qquad\qquad \qquad + \sum_{A=1}^N T^A S_A(q, \dot q, T^1,...,T^N),
\end{align*} 
where we assumed the function $(S_1,.., S_N) \in  \mathbb{R}  ^N  \mapsto \left( \frac{\partial L}{\partial S_1}, ..., \frac{\partial L}{\partial S_N} \right) \in  ( \mathbb{R}_+ ) ^N $ is a diffeomorphism for all $(q, \dot q) \in TQ_{ \boldsymbol{\Sigma} }$ and where the functions $S_A(q,\dot q, T^1,...,T^N)$, $A=1,...,N$ are defined from the conditions $ - \frac{\partial L}{\partial S_A}(q, \dot q, S_1,...,S_N)=T^A$, for all $A=1,...,N$ and for all $q, \dot q, S_1,...,S_N$. 
\medskip

Recall $T^{A}=\dot{\Gamma}^{A}$. We say that a curve $\left(q(t), \Gamma ^A (t) , \Sigma _A (t) \right) \in Q_{ \boldsymbol{\Sigma} } \times \mathbb{R}  ^{2N}  $, $t \in [t _1 , t _2 ]$ is a \textit{solution of the free energy variational formulation of nonequilibrium thermodynamics} if it satisfies the \textit{variational condition}
\begin{equation}\label{LdA_thermo_discrete_systems-new}
\begin{aligned} 
\delta \int_{ t _1 }^{ t _2 }\Big( \mathcal{L}(q  , \dot q , \dot \Gamma ^{1},...,\dot \Gamma ^{N} )&- \sum_{A=1}^N  \Sigma _A\dot\Gamma ^A \Big)   dt +\int_{ t _1 }^{ t _2 }\left\langle F^{\rm ext}, \delta q \right\rangle dt =0,
\end{aligned}
\end{equation}
for all variations $ \delta q , \delta \Gamma^A  , \delta \Sigma_A$ subject to the \textit{variational constraint}
\begin{equation}\label{Virtual_Constraints_Systems-new} 
\dot \Gamma  ^A  \delta \Sigma _A  = - \left\langle F^{{\rm fr}(A)}(...), \delta q  \right\rangle+ \sum_{B=1}^N \kappa _{AB}(\delta \Gamma  ^B-\delta \Gamma ^A ),\;\; \textrm{$($no sum on $A)$}
\end{equation}
with $ \delta q( t _i )=0$ and $ \delta \Gamma^A ( t _i )=0$ for $i=1,2$, and if it satisfies the \textit{phenomenological constraint}
\begin{equation}\label{Kinematic_Constraints_Systems-new} 
\dot \Gamma  ^A \dot{\Sigma}_A = -\left\langle F^{{\rm fr}(A)}(...), \dot q  \right\rangle + \sum_{B=1}^N \kappa _{AB}(\dot \Gamma  ^B- \dot \Gamma ^A ) + P_H^{{\rm ext} \rightarrow A}.
\end{equation} 
From this variational formulation, we deduce the system of evolution equations for the discrete system in terms of the free energy as

{\fontsize{9pt}{9pt}\selectfont
\begin{equation}\label{al_thermo_mech_equations_discrete_systems2}
\left\{ 
\begin{array}{l}
\displaystyle\vspace{0.1cm}\frac{d}{dt} \frac{\partial \mathcal{L}}{\partial \dot q }- \frac{\partial \mathcal{L}}{\partial q }= \sum_{A=1}^N F^{{\rm fr}(A)}+F^{\rm ext},\\[2mm]
\displaystyle T ^A \frac{d}{dt}\frac{\partial \mathcal{L}}{\partial T^{A}} = - \left\langle F^{{\rm fr}(A)} , \dot q \right\rangle +\sum_{B=1}^N \kappa _{AB}( T^{B}-T ^A )+P_H^{{\rm ext} \rightarrow A}, \quad A=1,...,N.
\end{array} 
\right.
\end{equation}}

\noindent For a free energy Lagrangian of the form $ \mathcal{L} (q, \dot q, T_1,...,T_N)= L_{\rm mech}(q, \dot q)- \sum_{A=1}^NF ^A (q, T_A)$, the \textit{heat equation} is computed from the second equation of \eqref{al_thermo_mech_equations_discrete_systems2} as
\[
c_{v,A} (q,T ^A) \dot T ^A=T^A \frac{\partial  ^2 F^A}{\partial q \partial T^A}(q,T^A)\dot q - \left\langle F^{\rm fr}, \dot q \right\rangle +\sum_{B=1}^N \kappa _{AB}( T^{B}-T ^A ) + P^{\rm ext}_H ,
\]
where $c_{v, A}(q,T^A )= - T^A \frac{\partial ^2 F^A}{\partial (T ^A)^2 }(q,T^A)$ is the specific heat of system $ \boldsymbol{\Sigma} _A$.

\section{The Navier-Stokes-Fourier equations}\label{3}

We shall now systematically extend  to the continuum setting the previous free energy variational formulations by focalising on the case of a heat conducting viscous fluid. We refer to \cite{GBYo2016b} for the corresponding formulation in terms of the entropy.

The variational formulation of the Navier-Stokes-Fourier equation is first formulated in the Lagrangian (or material) representation, because it is in this representation that the variational formulation is deduced from that of discrete systems described in \S\ref{2_3}. All the equations are intrinsically written in a differential-geometric form.

\subsection{Configuration space and Lagrangians}

We assume that the domain occupied by the fluid is a smooth compact manifold with smooth boundary $ \partial\mathcal{D}$. The configuration space is $Q= \operatorname{Diff}_0( \mathcal{D} )$, the group of all diffeomorphisms$^{3}$ of $ \mathcal{D} $ that keep the boundary $ \partial \mathcal{D} $ pointwise fixed. This corresponds to no-slip boundary conditions. This choice of the configuration space aims to describe only strong solutions of the partial differential equation. We assume  that the manifold $ \mathcal{D} $ is endowed with a Riemannian metric $g$.

\addtocounter{footnote}{1}
\footnotetext{In this paper we do not describe the functional analytic setting needed to rigorously work in the framework of infinite dimensional manifolds. For example, one can assume that the diffeomorphisms are of some given Sobolev class, regular enough (at least of class $C ^1$), so that $\operatorname{Diff}_0( \mathcal{D} )$ is a smooth infinite dimensional manifold and a topological group with smooth right translation, \cite{EbMa1970}. }

Given a curve $ \varphi _t$ of diffeomorphisms, starting at the identity at $t=0$, we denote by $ x= \varphi _t(X)= \varphi (t,X) \in \mathcal{D}$ the current position of a fluid particle which at time $t=0$ is at $X \in \mathcal{D} $. The \textit{mass density} $\varrho (t,X)$ and the \textit{entropy density} $S(t,X)$ in the Lagrangian (or material) description are respectively related to the corresponding quantities $ \rho (t,x)$ and $ s(t,x)$ in the Eulerian (or spatial) description as
\begin{equation}\label{material_rho_S} 
\varrho (t,X)= \rho (t, \varphi _t(X)) J _{\varphi _t}(X) \quad\text{and}\quad S(t,X)= s (t, \varphi_t (X)) J_{ \varphi _t}(X),
\end{equation} 
were $ J _{\varphi_t }$ denotes the Jacobian of $ \varphi _t$ relative to the Riemannian metric $g$, i.e., $ \varphi_t^\ast \mu _g = J_{\varphi_t} \mu _g $, with $ \mu _g $ the Riemannian volume form.

From the conservation of the total mass, we have $ \varrho (t, X)= \varrho  _{\rm ref}(X)$, i.e., the mass density in the material description is time independent. It therefore appears as a parameter in the Lagrangian function and in the variational formulation. This is not the case for the material entropy $S(t,X)$, which is a dynamic field with corresponding variations $ \delta S$ that must be taken into account in the variational formulation.

\medskip

\noindent \textbf{The Lagrangian.} In a similar way to the case of discrete systems in \eqref{Lagrangian_S_discrete}, the Lagrangian in the material description is a map
\begin{equation}\label{Lagrangian} 
L_{ \varrho _{\rm ref}}: T \operatorname{Diff}_0( \mathcal{D} ) \times \mathcal{F} ( \mathcal{D} )\rightarrow \mathbb{R}, \quad ( \varphi , \dot \varphi , S) \mapsto L_{ \varrho _{\rm ref}}( \varphi , \dot \varphi , S),
\end{equation} 
where $T \operatorname{Diff}_0( \mathcal{D} )$ is the tangent bundle to $ \operatorname{Diff}_0( \mathcal{D} )$ and $ \mathcal{F} ( \mathcal{D} )$ is a space of real valued functions on $ \mathcal{D} $ with a given high enough regularity, so that all the formulas used below are valid. The index notation in $L_{\rm \varrho _{\rm ref}}$ is used to recall that $L$ depends parametrically on $ \varrho _{\rm ref}$. By $( \varphi , \dot \varphi )$ we denote an arbitrary vector in the tangent space $T_ \varphi \operatorname{Diff}_0( \mathcal{D} )$. We choose to follow here the traditional coordinate notation$^{4}$ \addtocounter{footnote}{1}
\footnotetext{The intrinsic geometric notation is $L_{ \varrho _{\rm ref}}( \mathbf{V} _ \varphi , S)$ for $ \mathbf{V} _ \varphi \in T _\varphi \operatorname{Diff}_0( \mathcal{D} )$.}(i.e., of the type $L(q, \dot q)$) used in classical mechanics and in \S\ref{section_2}, although our point of view is completely intrinsic. Recall that the tangent space to $\operatorname{Diff}_0( \mathcal{D} )$ at $ \varphi $  is given by $T_ \varphi \operatorname{Diff}_0(\mathcal{D} )= \{ \mathbf{V} _\varphi : \mathcal{D} \rightarrow T \mathcal{D} \mid \mathbf{V} _ \varphi (X) \in T_{\varphi (X)} \mathcal{D},\;\; \mathbf{V} _\varphi |_ {\partial \mathcal{D}} =0 \}$, where the map $ \mathbf{V} _\varphi  : \mathcal{D} \rightarrow T \mathcal{D}$   has the same regularity with $ \varphi $.

Consider a gas with a given state equation $ \varepsilon = \varepsilon (\rho, s)$ where $ \varepsilon $ is the internal energy density, and the Lagrangian is given by

{\fontsize{9pt}{9pt}\selectfont
\begin{equation}\label{Lagrangian_NSF}
\begin{aligned} 
L_{ \varrho _{\rm ref}}( \varphi , \dot\varphi , S)&\!=\! \int_ \mathcal{D}  \frac{1}{2}\varrho  _{\rm ref}( X ) | \dot\varphi ( X )| ^2 _g \mu _g(X) \!-\! \int_ \mathcal{D} \varepsilon \left( \frac{ \varrho  _{\rm ref}(X)}{J_\varphi (X)} ,  \frac{ S(X)}{J_\varphi (X)} \right) J_\varphi (X) \mu _g(X)\\
&= \int_ \mathcal{D} \mathfrak{L} (\varphi (X)  , \dot \varphi (X),T _X\varphi , \varrho _{\rm ref}(X),S(X)) \mu _g (X), 
\end{aligned}
\end{equation}}

\noindent where $T _ X\varphi :T_ X \mathcal{D} \rightarrow T_ { \varphi (X)}\mathcal{D} $ is the tangent map to $ \varphi $. The first term of $L_{ \varrho _{\rm ref}}$ represents the total kinetic energy of the gas, computed with the help of the Riemannian metric $g$, and the second term represents the total internal energy. The second term is deduced from $\varepsilon ( \rho , s)$ by using the relations \eqref{material_rho_S}. Both terms are written here in terms of material quantities. In the second line we defined the Lagrangian density $\mathfrak{L}  (\varphi   ,\dot \varphi ,  T \varphi , \varrho _{\rm ref},S)\mu _g$ as the integrand of the Lagrangian $L$. The \textit{material temperature} is given by
\[
\mathfrak{T}= - \frac{\partial \mathfrak{L}  }{\partial S}= \frac{\partial \varepsilon  }{\partial s}( \rho , s)  \circ \varphi = T \circ\varphi ,
\]
where $T$ is the \textit{Eulerian temperature}.

\medskip

\noindent \textbf{The free energy Lagrangian.} 
Generally, given a Lagrangian $L_{\varrho _{\rm ref}}:T \operatorname{Diff}_0( \mathcal{D} ) \times \mathcal{F} ( \mathcal{D} )\rightarrow \mathbb{R}$ with Lagrangian density $ \mathfrak{L} (\varphi   ,\dot \varphi ,  T \varphi , \varrho _{\rm ref},S)$, we define the associated \textit{free energy Lagrangian} $\mathcal{L} _{\varrho _{\rm ref}}: T \operatorname{Diff}_0( \mathcal{D} ) \times \mathcal{F} _+ ( \mathcal{D} )\rightarrow \mathbb{R}$ as 
\begin{equation}\label{FEL}
\begin{aligned} 
&\mathcal{L}_{\varrho _{\rm ref}} ( \varphi , \dot\varphi , \mathfrak{T}  ):=\int_ \mathcal{D} \mathscr{L}(\varphi (X), \dot\varphi (X),T_X\varphi ,  \varrho _{\rm ref}(X),\mathfrak{T}  (X)) \mu _g,
\end{aligned}
\end{equation} 
where  the free energy Lagrangian density $\mathscr{L}$ is defined by
\begin{equation}\label{definition_mathscrL} 
\mathscr{L}( \alpha ,\mathfrak{T}  ):= \mathfrak{L} (\alpha , S( \alpha , \mathfrak{T} )) + \mathfrak{T}  S(\alpha ,\mathfrak{T} ).
\end{equation} 
In \eqref{definition_mathscrL} we used the abbreviation $ \alpha = (\varphi , \dot\varphi ,T \varphi , \varrho _{\rm ref})$, we assumed that the function $ S \in\mathbb{R} \mapsto \frac{\partial\mathfrak{L}  }{\partial S}( \alpha , S) \in \mathbb{R}  _+ $ is invertible for all $ \alpha $, and we defined the function $ S( \alpha , \mathfrak{T} )$ by the condition $ \frac{\partial \mathcal{L} }{\partial S}(\alpha , S( \alpha , \mathfrak{T}  ) )= \mathfrak{T}  $, for all $ \alpha $. 
In \eqref{FEL}, $\mathcal{F} _+ ( \mathcal{D} )$ denotes a set of strictly positive functions on $ \mathbb{R}  $.

For the case of the Lagrangian \eqref{Lagrangian_NSF}, we obtain

{\fontsize{9pt}{9pt}\selectfont
\begin{equation}\label{FE_Lagrangian_NSF}
\begin{aligned} 
\mathcal{L} _{\varrho _{\rm ref}}( \varphi , \dot\varphi , & \mathfrak{T}  )=: \int_ \mathcal{D} \mathscr{L} \big( \varphi (X),\dot \varphi (X), T _X\varphi , \varrho  _{\rm ref}( X ), \mathfrak{T}  (X)\big)\mu _g(X)\\
&= \int_ \mathcal{D} \frac{1}{2}\varrho  _{\rm ref}( X ) | \dot\varphi ( X )| ^2 _g \mu _g(X) - \int_ \mathcal{D} \psi  \left( \frac{ \varrho  _{\rm ref}(X)}{J_\varphi (X)} , \mathfrak{T}  (X)\right) J_\varphi (X) \mu _g(X),
\end{aligned}
\end{equation}}

\noindent where $ \psi ( \rho , T)$ is the  (Helmholtz) free energy density associated to the internal energy density $ \varepsilon ( \rho , s)$. The free energy density is given in material representation as
\begin{equation}\label{FE_material} 
\Psi ( \varphi , T_X \varphi , \varrho _{\rm ref}, \mathfrak{T}  ):= \psi  \left( \frac{ \varrho  _{\rm ref} }{J_\varphi } , \mathfrak{T}   \right) J_\varphi.
\end{equation}

For example, for the case of a perfect gas, the internal energy density and the free energy density are respectively given by
\begin{align*} 
\varepsilon ( \rho , s)&= \varepsilon _0 e^ {\frac{1}{C_v} \left( \frac{s}{ \rho }-\frac{s_0}{\rho _0}   \right) }\left( \frac{ \rho }{\rho _0}\right)  ^{C_p/C_v},\\
\psi ( \rho ,T)&=\rho T \left(\frac{\psi _0 }{\rho _0 T_0} +  R _{\rm gas} \ln \left( \frac{ \rho }{\rho _0}  \right) -C_v \ln \left( \frac{T}{T_0}\right) \right),
\end{align*} 
where the constant $C_v$ is the specific heat coefficient at constant volume, the constant $C_p$ is the specific heat at constant pressure, $R_{\rm gas}= C_p-C_v$ is the gas constant, and $ \rho _0, s_0, T_0, \varepsilon _0, \psi _0$ are given constant reference values verifying $ \varepsilon _0= \varepsilon ( \rho _0, s_0)$ and $ \psi _0= \psi ( \rho _0, T_0)$.

\subsection{Variational formulation in material representation}\label{3_2} 

It is well-known that in absence of irreversible processes the equations of continuum mechanics in material representation arise from Hamilton's principle, see, e.g., \cite[Ch.5]{MaHu1983}. Before presenting the extension of this variational formulation to the irreversible case, we shall briefly review Hamilton's principle as it applies to the Lagrangian 
\eqref{Lagrangian_NSF}. In this case, the resulting system becomes the \textit{perfect adiabatic compressible fluid}.

\medskip

\noindent \textbf{Hamilton's principle in the reversible case.} 
In absence of irreversible processes, the entropy is conserved so that we have $S(t, X)= S_{\rm ref}(X)$ in material representation. The equations of motion thus follow from Hamilton's principle applied to the Lagrangian $L_{\varrho _{\rm ref}}$ in \eqref{Lagrangian_NSF}, in which the variable $S=S_{\rm ref}$ is understood as a fixed parameter field  in the same way as $\varrho _{\rm ref}$. Writing $L_{\varrho _{\rm ref}, S_{\rm ref}}:T \operatorname{Diff}_0( \mathcal{D} ) \rightarrow \mathbb{R}$, with $L_{\varrho _{\rm ref}, S_{\rm ref}}( \varphi , \dot \varphi ):= L_{\varrho _{\rm ref}}(\varphi , \dot\varphi , S_{\rm ref})$, Hamilton's principle is
\begin{equation}\label{HP_easy} 
\delta \int_ {t_1}^{t_2} L_{\varrho _{\rm ref}, S_{\rm ref}}( \varphi , \dot\varphi )dt=0 ,
\end{equation} 
with respect to variations $ \delta \varphi $ such that $\delta \varphi |_{ \partial\mathcal{D} }=0$ and $ \delta \varphi (t _i )=0$, $i=1,2$. The stationarity condition yields the Euler-Lagrange equations for $L_{\varrho _{\rm ref}, S_{\rm ref}}$ in the form
\begin{equation}\label{EL_fluids} 
\varrho _{\rm ref}\frac{D}{Dt} \mathbf{V} = \operatorname{DIV} \mathbf{P} ^{\rm cons}, 
\end{equation} 
where $ \mathbf{V} $ is the material velocity and $ \mathbf{P} ^{\rm cons}$ is the first Piola-Kirchhoff stress tensor.

We now describe in details the geometric objects involved in this equation. The material velocity is defined by $ \mathbf{V}_t(X):= \frac{\partial }{\partial t} \varphi _ t(X) \in T_{ \varphi_t(X)}\mathcal{D} $. At each time $t$, the material velocity $ \mathbf{V} _t$ is a vector field on $ \mathcal{D} $ along $ \varphi_t$. The first Piola-Kirchhoff tensor is defined by
\begin{equation}\label{Pcons1} 
\mathbf{P} ^{\rm cons}= - \left( \frac{\partial \mathfrak{L} }{\partial T_{X} \varphi } \right) ^\sharp,
\end{equation}
where $ \sharp$ denotes the index rising operator relative to the Riemannian metric $g$.
At each time $t$, the first Piola-Kirchhoff tensor $ \mathbf{P}  ^{\rm cons}(t,\_\,)$ is a two-point tensor along $ \varphi _t$. More precisely, for all $X \in \mathcal{D} $, we have
\[
\mathbf{P} ^{\rm cons}(t,X): T_ x ^\ast \mathcal{D} \times T_ X ^\ast \mathcal{D} \rightarrow \mathbb{R}  , \quad \text{where} \quad  x= \varphi _t(X).
\]
On the left hand side of \eqref{EL_fluids}, $D \mathbf{V} /Dt$ denotes the covariant time derivative of the vector field $ \mathbf{V} _t$ along $ \varphi_t $, relative to the Riemannian metric $g$. This covariant derivative yields again a vector field along $ \varphi _t$. On the right hand side of \eqref{EL_fluids}, $ \operatorname{DIV} \mathbf{P} ^{\rm cons}$ denotes the divergence${}^4$ of the two-point tensor $ \mathbf{P} ^{\rm cons}$ along $ \varphi _t $, relative to the Riemannian metric $g$. It yields a vector field on $ \mathcal{D} $ along $ \varphi _t $. We refer to \cite{MaHu1983} for a detailed account of two-point tensors along diffeomorphisms and their covariant derivatives.


A direct computation shows that for the Lagrangian \eqref{Lagrangian_NSF}, the first Piola-Kirchhoff tensor leads to the pressure in material representation, i.e.,  
\begin{equation}\label{P_cons_NSF} 
\mathbf{P} ^{\rm cons}(X)( \alpha _x, \beta _X)= - p ( \rho , s) J _\varphi \,g\big( T_x\varphi^{-1} ( \alpha _x), \beta _X\big), \quad p( \rho , s)= \frac{\partial \varepsilon}{\partial \rho }\rho + \frac{\partial \varepsilon }{\partial s}s - \varepsilon,
\end{equation} 
where $x= \varphi _t(X)$.

The system \eqref{EL_fluids}, with $ \mathbf{P} ^{\rm cons}$ given in \eqref{P_cons_NSF},  is the material description of the Euler equations for a compressible perfect adiabatic fluid, which is usually written in spatial representation as
\begin{equation}\label{viscoelastic_spat}
\left\{
\begin{array}{l}
\vspace{0.2cm}\rho ( \partial _t \mathbf{v} + \mathbf{v}  \cdot \nabla  \mathbf{v} )=- \operatorname{grad}  p,\\[1mm]
\partial _t \rho + \operatorname{div} ( \rho \mathbf{v} )=0, \quad \partial _t s+ \operatorname{div}(s \mathbf{v} )=0,
\end{array}
\right.
\end{equation}
where $ \nabla $ is the Levi-Civita covariant derivative associated to $g$ and the operators $ \operatorname{grad}$ and $ \operatorname{div}$ are both associated to $g$. A systematic approach to derive the variational principle in spatial representation from Hamilton's principle in material representation, is provided by the Euler-Poincar\'e reduction theory on Lie groups, see \cite{HoMaRa1998}. 

\begin{remark}[Free energy Lagrangian in the reversible case]{\rm It is important to observe that  while the entropy variable $S_{\rm ref}(X)$ is seen as a fixed parameter in the material description of the reversible case, this is not true for the temperature as it clearly follows from its definition $ \mathfrak{T}  = - \frac{\partial \mathfrak{L} }{\partial S}(\varphi , \dot\varphi ,T \varphi , \varrho _{\rm ref}, S_{\rm ref})$. Therefore, when working with the corresponding free-energy Lagrangian $ \mathcal{L}_{ \varrho _{\rm ref}} ( \varphi , \dot\varphi , \mathfrak{T}  )$ in \eqref{FE_Lagrangian_NSF}, one cannot avoid considering the variations $ \delta \mathfrak{T}  $ in the variational principle, even in the reversible case. We will consider this situation later, as a particular case of our variational formulation for the irreversible case.}
\end{remark} 

\begin{remark}\label{Material_G}{\rm The general geometric formulation of continuum mechanics needs a more general setting than the one presented above, namely, one has to consider the field $ \varphi $ as an embedding from a reference manifold $ \mathcal{B}$  into an ambient manifold $ \mathcal{S} $. Each of these manifolds is endowed with a Riemannian metric, typically denoted by $ G$ on $ \mathcal{B} $ and by $g$ on $ \mathcal{S} $. This is the appropriate setting to study the material and spatial symmetries in nonlinear continuum mechanics, see \cite{MaHu1983}, \cite{SiMaKr1988}, \cite{GBMaRa2012}. In the present situation, since we are studying the special case of a fluid in a fixed domain, we have $ \mathcal{B} = \mathcal{S} =\mathcal{D}$, so we can make the choice $g=G$.}
\end{remark}


\noindent \textbf{Navier-Stokes-Fourier equations in material representation.} The processes of viscosity and heat conduction are described by the inclusion of the corresponding thermodynamic fluxes given, in the material description, by a friction Piola-Kirchhoff tensor $ \mathbf{P} ^{\rm fr} (t, X)$ and an entropy flux density $ \mathbf{J} _S(t, X)$, where $ \mathbf{P} ^{\rm fr}$ is a two-point tensor along $ \varphi $ and $ \mathbf{J} _S$ is a vector field on $ \mathcal{D} $. We will recall their usual phenomenological expressions later in the Eulerian description.

By complete analogy with the case of discrete systems developed earlier, our formulation needs the notion of \textit{thermal displacements} (\cite{GrNa1991}) in material representation, i.e., a variable $ \Gamma (t, X)$ such that
\[
\frac{d}{dt} \Gamma (t, X)=\mathfrak{T}  (t, X).
\]
This is a particular case of \textit{thermodynamic displacement variables} introduced in \cite{GBYo2016a,GBYo2016b}. 
In addition to these internal irreversible processes, we assume that the fluid is heated from the exterior, which is represented by a heat power supply density $ \varrho  _{\rm ref}(X)R(t,X)$ in material representation. Recall that $ \mathscr{L}$ denotes the free energy Lagrangian density.

\medskip

By complete analogy with the variational formulation given in \eqref{LdA_thermo_discrete_systems-new}--\eqref{Kinematic_Constraints_Systems-new} for discrete systems, we consider the variational formulation for a curve $( \varphi (t), \Gamma (t), \Sigma (t)) \in \operatorname{Diff}_0( \mathcal{D} ) \times \mathcal{F} ( \mathcal{D} ) \times \mathcal{F} ( \mathcal{D} ) $ as follows:
\begin{equation}\label{VC_NSF_material} 
\delta \int_{t_1}^{t_2} \!\!\int_ \mathcal{D} \left[  \mathscr{L} \big( \varphi , \dot \varphi , T\varphi , \varrho  _{\rm ref}, \dot \Gamma\big)- \Sigma \dot \Gamma\right]  \mu _g dt=0, \quad\;\;\; \textsc{\small Variational Condition}
\end{equation} 
for all variations $\delta\varphi$, $ \delta \Gamma $, $ \delta \Sigma $ subject to the constraint
\begin{equation}\label{CV_NSF_material} 
\dot\Gamma \delta \Sigma = (\mathbf{P} ^{\rm fr})^\flat : \nabla \delta \varphi -\mathbf{J} _S \cdot \mathbf{d} \delta \Gamma, \qquad\, \textsc{\small Variational Constraint}
\end{equation} 
with $ \delta \varphi (t_i)=0$ and $ \delta \Gamma (t_i)=0$, for $i=1,2$, where the curve $( \varphi (t), \Gamma (t), \Sigma (t))$ satisfies the constraint
\begin{equation}\label{KC_NSF_material}
\dot\Gamma \dot \Sigma = (\mathbf{P} ^{\rm fr}) ^\flat : \nabla \dot \varphi -\mathbf{J} _S \cdot \mathbf{d} \dot\Gamma+ \varrho  _{\rm ref}R, \qquad\, \textsc{\small Phenomenological Constraint}.
\end{equation} 

\medskip

In the constraints \eqref{CV_NSF_material} and \eqref{KC_NSF_material}, $ \delta \varphi , \dot\varphi : \mathcal{D} \rightarrow T \mathcal{D} $ are vector fields on $ \mathcal{D} $ covering $ \varphi $. The objects $ \nabla  \delta \varphi $ and $ \nabla \dot\varphi$ are two-point tensor fields covering $ \varphi $, obtained by taking the Levi-Civita covariant derivative associated to $g$. The notation $``:"$ means the full contraction of the two-point tensor fields along $ \varphi $. The flat operator on $ \mathbf{P} ^{\rm cons}$ applies to the spatial index.

\medskip

Note the analogy between the three conditions \eqref{LdA_thermo_discrete_systems-new}-\eqref{Virtual_Constraints_Systems-new}-\eqref{Kinematic_Constraints_Systems-new} for the variational formulation of discrete systems and the three conditions \eqref{VC_NSF_material}-\eqref{CV_NSF_material}-\eqref{KC_NSF_material} above. As before, the constraint \eqref{KC_NSF_material} is nonlinear and one can pass from the variational constraint to the phenomenological constraint by formally replacing the variations $ \delta  \varphi $, $\delta \Gamma $, $ \delta \Sigma $ by the time derivatives $ \dot\varphi $, $\dot \Gamma $, $\dot\Sigma $. 
This variational formulation does not impose any restriction concerning the dependance of the thermodynamic fluxes on the variables $ \varphi , \dot \varphi , \varrho _{\rm ref}, \Gamma , \dot \Gamma $ and on their spatial derivatives.

\medskip

Since $ \delta \varphi |_{ \partial \mathcal{D} }=0$ and $ \delta \varphi (t _i )=0$, $ \delta \Gamma (t _i )=0$, $i=1,2$, by applying the variational condition \eqref{VC_NSF_material}, we get
\[
\int_{t _1 }^{ t _2 } \!\!\int_ \mathcal{D} \left[ \left( \frac{\partial^ \nabla  \mathscr{L} }{\partial \varphi }- \operatorname{DIV} \frac{\partial \mathscr{L} }{\partial T_X \varphi }- \frac{D}{Dt}\frac{\partial \mathscr{L} }{\partial \dot \varphi }\right) \delta \varphi  + \left( \dot \Sigma - \frac{d}{dt}\frac{\partial \mathscr{L}}{\partial\dot  \Gamma }  \right)  \delta \Gamma  -  \dot \Gamma \delta \Sigma \right] \mu _g dt=0.
\]
Here $ \frac{\partial^ \nabla  \mathscr{L} }{\partial \varphi }$ is the partial derivative of $\mathscr{L} $ relative to the spatial coordinate $x= \varphi (X)$. Such a partial derivative needs to be defined with respect to a Riemannian metric, here $g$, this is why we insert the exponent $ \nabla $. For the Lagrangian in \eqref{FE_Lagrangian_NSF}, we have  $ \frac{\partial^ \nabla  \mathscr{L} }{\partial \varphi }=0$. The other partial derivatives of $ \mathscr{L}$ do not need the use of a Riemannian metric. The operator $ D/Dt$ and $ \operatorname{DIV}$ are associated to $g$, as explained earlier. 
Using the variational constraint \eqref{CV_NSF_material} and again $ \delta \varphi |_{ \partial \mathcal{D} }=0$ and $ \delta \varphi (t _i )=0$, $ \delta \Gamma (t _i )=0$, $i=1,2$, and collecting the terms associated to the variations $ \delta \varphi $ and $ \delta \Gamma $, we obtain the system
\begin{align} 
\delta \varphi : \quad &\;\;\rho _{\rm ref} \frac{D \mathbf{V} }{Dt}= \operatorname{DIV} \left(  \mathbf{P} ^{\rm cons}+\mathbf{P} ^{\rm fr} \right) , \label{delta_phi} \\
\delta \Gamma  : \quad &\;\; \frac{d}{dt}\frac{\partial \Psi }{\partial \mathfrak{T}  }  = \operatorname{DIV}\mathbf{J} _S  -\dot \Sigma\quad\text{and}\quad \mathbf{J} _S \cdot \mathbf{n} ^\flat =0 \;\; \text{on $ \partial\mathcal{D} $}, \label{delta_Gamma}
\end{align}
where $ \mathbf{n} $ is the outward pointing unit normal vector field along the boundary $ \partial \mathcal{D} $ and $ \mathbf{n} ^\flat $ is the associated one-form which is obtained by applying the index lowering operator $ \flat $ associated to $g$. The second equation in \eqref{delta_Gamma} arises from the freeness of the variations $ \delta \Gamma $ at the boundary. In \eqref{delta_phi} the conservative Piola-Kirchhoff stress tensor $ \mathbf{P} ^{\rm cons}$ is defined from the free energy Lagrangian density $ \mathscr{L}$ (or the free energy density $\Psi $) as
\[
 \mathbf{P} ^{\rm cons}= - \left( \frac{\partial \mathscr{L} }{\partial T_{X} \varphi } \right) ^\sharp=\left( \frac{\partial \Psi  }{\partial T_{X} \varphi } \right) ^\sharp.
\]
Here we note that $ \mathbf{P} ^{\rm cons}$ is defined in terms of $ \mathscr{L}$ whereas in \eqref{Pcons1} it is defined in terms of $\mathfrak{L}$. From the general definition of the free energy Lagrangian density in \eqref{definition_mathscrL}, we see that these two definitions of $ \mathbf{P} ^{\rm cons}$ coincide, one being expressed in terms of the entropy \eqref{Pcons1}, and the other with respect to the temperature. For the free energy Lagrangian  \eqref{FE_Lagrangian_NSF}, we can compute the partial derivative with respect to $ T_ X \varphi $ to get
\[
\mathbf{P} ^{\rm cons}( \alpha _x, \beta _X)= - p ( \rho , T) J_ \varphi \,g\big( T_x\varphi^{-1} ( \alpha _x), \beta _X\big), \quad p( \rho , T)= \frac{\partial \psi }{\partial \rho }\rho - \psi ,
\]
where $x= \varphi_{t} (X)$. This expression coincides with \eqref{P_cons_NSF} although it is here expressed in terms of the temperature.

Using the relation \eqref{delta_Gamma} in the phenomenological constraint \eqref{KC_NSF_material} yields 
\[
\mathfrak{T}  \left( -\frac{d}{dt}\frac{\partial \Psi }{\partial \mathfrak{T}  } +  \operatorname{DIV} \mathbf{J} _S\right)   = ( \mathbf{P} ^{\rm fr})^\flat: \nabla  \dot \varphi - \mathbf{J} _S \cdot \mathbf{d}  \mathfrak{T}  + \varrho  _{\rm ref}R,
\]
where we recall that $\mathfrak{T}  :=  \dot \Gamma $ is the temperature in material representation.
\medskip

These results are summarized as follows.

\begin{theorem}[Free energy variational formulation for the Navier-Stokes-Fourier system -- material representation]
The Navier-Stokes-Fourier equations in material representation, given by
\begin{equation}\label{summary_NSF} 
\left\{
\begin{array}{l}
\vspace{0.2cm}\displaystyle\varrho _{\rm ref} \frac{D \mathbf{V} }{Dt}= \operatorname{DIV}( \mathbf{P} ^{\rm cons}+ \mathbf{P} ^{\rm fr}),\\[2mm]
\displaystyle\mathfrak{T}  \left( -\frac{d}{dt}\frac{\partial \Psi }{\partial \mathfrak{T}  }+  \operatorname{DIV} \mathbf{J} _S \right)   = ( \mathbf{P} ^{\rm fr}) ^\flat : \nabla  \mathbf{V}  - \mathbf{J} _S \cdot \mathbf{d} \mathfrak{T}  + \varrho  _{\rm ref}R,
\end{array}
\right.
\end{equation}
with boundary conditions $\mathbf{V} |_{ \partial \mathcal{D} }=0$ and $\mathbf{J} _S \cdot \mathbf{n} ^\flat |_{ \partial \mathcal{D} }= 0 $, follow from the variational condition \eqref{VC_NSF_material} together with the variational and phenomenological constraints \eqref{CV_NSF_material} and \eqref{KC_NSF_material}.
\end{theorem}

When applied to a general free energy Lagrangian density $ \mathscr{L}$, this theorem yields the general system
\begin{equation}\label{general_L} 
\left\{
\begin{array}{l}
\vspace{0.2cm}\displaystyle \frac{D }{Dt} \frac{\partial  \mathscr{L}}{\partial \dot \varphi } - \frac{\partial ^{\nabla\!\!}\mathscr{L}}{\partial \varphi } = \operatorname{DIV} \left( - \frac{\partial \mathscr{L}}{\partial T_ X \varphi } + (\mathbf{P} ^{\rm fr} ) ^\flat \right) ,\\[2mm]
\displaystyle\mathfrak{T}  \left(  \frac{d}{dt} \frac{\partial \mathscr{L}}{\partial \mathfrak{T}} +  \operatorname{DIV} \mathbf{J} _S \right)   = ( \mathbf{P} ^{\rm fr}) ^\flat : \nabla  \mathbf{V}  - \mathbf{J} _S \cdot \mathbf{d} \mathfrak{T}  + \varrho  _{\rm ref}R,
\end{array}
\right.
\end{equation}
with boundary conditions $\mathbf{V} |_{ \partial \mathcal{D} }=0$ and $\mathbf{J} _S \cdot \mathbf{n} ^\flat |_{ \partial \mathcal{D} }= 0 $. 
This system is the continuum version of system \eqref{al_thermo_mech_equations_discrete_systems2} and describes the equations of motion for a continuum theory with free energy Lagrangian density $\mathscr{L}$ and subject to the irreversible processes of viscosity and heat conduction.

\medskip

\noindent \textbf{Geometric structure associated to the variational formulation.} We now comment on the general geometric setting underlying the formulation \eqref{VC_NSF_material}--\eqref{KC_NSF_material} and its relation with the variational formulation in nonholonomic mechanics. Let us consider formally the manifold $ \mathcal{Q} :=  \operatorname{Diff}_0( \mathcal{D} ) \times \mathcal{F} ( \mathcal{D} ) \times \mathcal{F} ( \mathcal{D} )$ and denote an element in $\mathcal{Q}$ by $q:= (\varphi , \Gamma , \Sigma)$.
We consider the vector bundle $T  \mathcal{Q} \times _\mathcal{Q} T \mathcal{Q} \rightarrow \mathcal{Q} $ whose vector fiber at $q \in \mathcal{Q} $ is given by the vector space $T_q \mathcal{Q} \times T_q \mathcal{Q} $. For convenience, we shall write an element in this fiber by using the local$^{5}$ notation $( q, \dot q, \delta q) \in T_q \mathcal{Q} \times T_q \mathcal{Q}$.

\addtocounter{footnote}{1}
\footnotetext{The intrinsic notation of such an element is $(v_q, w_q) \in T_qQ \times T_qQ$, where $v_q, w_q \in T_qQ$.}

Geometrically, the variational constraint \eqref{CV_NSF_material} defines a subset $C_V \subset T  \mathcal{Q}  \times _ \mathcal{Q} T \mathcal{Q}$ as follows: $( q, \dot q, \delta q) \in C_V$ $\Leftrightarrow$ $( q, \dot q, \delta q)$ satisfies \eqref{CV_NSF_material},
where $( q, \dot q, \delta q)= (\varphi , \Gamma , \Sigma,\dot \varphi , \dot\Gamma , \dot\Sigma, \delta \varphi ,  \delta \Gamma , \delta \Sigma)$.
This variational constraint satisfies the following property: for each $(q, \dot q) \in T \mathcal{Q} $, the set $C_V(q, \dot q)$ defined by $C_V(q, \dot q):= C_V \cap \{(q, \dot q)\} \times T_q \mathcal{Q}$ is a vector space.

The phenomenological constraint \eqref{KC_NSF_material} on $(q, \dot q)= (\varphi , \Gamma , \Sigma,\dot \varphi , \dot\Gamma , \dot\Sigma)$ geometrically defines a subset $C_K \subset T \mathcal{Q}$ of the tangent bundle to $\mathcal{Q} $.
For the case of adiabatically closed systems (i.e., $ \varrho  _{\rm ref} R=0$), the subset $C_K$ is obtained from the variational constraint $C_V$ via the following general construction
\begin{equation}\label{CKCV} 
C_K:=\{(q, \dot q) \in T \mathcal{Q} \mid (q, \dot q)\in C_V(q, \dot q)\}.
\end{equation} 
Constraints $C_K$ and $C_V$ are related as in \eqref{CKCV}, which we refer to as \textit{constraints of the thermodynamic type}, see \cite{GBYo2017c}.

In terms of the above constraint sets $C_V$ and $C_K$, the variational formulation \eqref{VC_NSF_material}--\eqref{KC_NSF_material} of the Navier-Stokes-Fourier system can thus be written as follows:

A curve $q(t) = ( \varphi (t), \Gamma (t), \Sigma (t))\in \mathcal{Q} $, $t \in [t _1 , t _2 ]$ satisfies the Navier-Stokes-Fourier system \eqref{summary_NSF} if and only if it satisfies the variational condition
\begin{equation}\label{VC_general} 
\delta \int_{t_1}^{t_2} \mathsf{L}(q, \dot q) dt=0,
\end{equation} 
for all variations
\begin{equation}\label{CV_general} 
\delta q \in C_V(q, \dot q)
\end{equation} 
with $ \delta q(t _1 )= \delta q(t _2)=0$ and where the curve $q(t)$ satisfies
\begin{equation}\label{CK_general} 
\dot q(t) \in C_K.
\end{equation} 
The Lagrangian $\mathsf{L}$ in \eqref{VC_general} denotes the full expression under the time integral in \eqref{VC_NSF_material}, namely,
\[
\mathsf{L}(q,\dot{q})=\int_ \mathcal{D} \left[  \mathscr{L} \big( \varphi , \dot \varphi , T\varphi , \varrho  _{\rm ref}, \dot \Gamma\big)- \Sigma \dot \Gamma\right]  \mu _g.
\]

From a mathematical point of view, the variational formulation of \eqref{VC_general}--\eqref{CK_general} is a {\it nonlinear} (and infinite dimensional) extension of the Lagrange-d'Alembert principle used for the treatment of nonholonomic mechanical systems with \textit{linear} constraints, see e.g., \cite{Bl2003}. Such linear constraints are given by a distribution $ \Delta  \subset T \mathcal{Q} $ on $ \mathcal{Q} $. In this linear case, we have $C_K= \Delta \subset T \mathcal{Q} $ and the variational constraint is $C_V= T \mathcal{Q} \times _Q \Delta $.

For the case of \textit{nonlinear} constraints $C _K\subset T  \mathcal{Q} $ on velocities in \textit{mechanics}, which are called {\it kinematic constraints}, a generalization of the Lagrange-d'Alembert principle has been considered in \cite{Ch1934}, see also \cite{Ap1911}, \cite{Pi1983}. In Chetaev's approach, the variational constraint $C_{V}$ is derived from the kinematic constraint $C_K$. However, it has been pointed out in \cite{Ma1998} that this principle does not always lead to the correct equations of motion for mechanical systems and in general one has to consider the kinematic and variational constraints as independent notions. A general geometric variational approach for nonholonomic systems with nonlinear and (possibly) higher order kinematic and variational constraints has been described in \cite{CeIbdLdD2004}. This setting generalizes both the Lagrange-d'Alembert and Chetaev approaches.
It is important to point out that for these generalizations, including Chetaev's approach, energy may not be conserved along the solution of the equations of motion. The variational formulation \eqref{VC_general}--\eqref{CK_general} falls into the general setting described in \cite{CeIbdLdD2004}, extended here to the infinite dimensional setting.
In the special case of constraints of the thermodynamic type, i.e., related through \eqref{CKCV}, the energy is conserved, see \cite{GBYo2017c}, consistently with the fact that in such a situation the system is isolated.

\subsection{Variational formulation in spatial representation}

We shall now develop the spatial (or Eulerian) representation of the variational formulation \eqref{VC_NSF_material}-\eqref{VC_NSF_material}-\eqref{KC_NSF_material}. The spatial  fields associated to $\dot \varphi $, $ \varrho _{\rm ref}$, $\Gamma $, $ \Sigma $, $ \mathfrak{T}  $, are denoted by $\mathbf{v}$, $ \rho $, $s$, $ \sigma $, $ \gamma $, $T$. The spatial and Lagrangian fields are related as follows
\begin{equation}\label{def_Eulerian_variables} 
\begin{array}{lll} 
\vspace{0.2cm}&\mathbf{v}= \dot \varphi\circ\varphi ^{-1},  &\rho =(\varrho _{\rm ref}\circ \varphi^{-1} )J_ \varphi ^{-1},\\
\vspace{0.2cm}&s = (S\circ \varphi^{-1} )J_ \varphi ^{-1},  \qquad & \sigma  = (\Sigma \circ \varphi^{-1} )J_ \varphi ^{-1},\\
\vspace{0.2cm}&\gamma = \Gamma \circ \varphi^{-1} , & T=\mathfrak{T} \circ \varphi^{-1}.
\end{array}
\end{equation} 
The Eulerian quantities associated to $ \mathbf{J} _S$, $ \mathbf{P} ^{\rm fr}$, $R$ are denoted by $ \mathbf{j} _s$, $ \boldsymbol{\sigma} ^{\rm ref}$, $r$, and are defined as
\begin{equation}\label{def_Eulerian_flux}
\begin{aligned}
&\mathbf{j}_s= (T \varphi \circ  \mathbf{J} _S \circ \varphi^{-1}  )J_\varphi ^{-1}, \qquad r= R \circ \varphi^{-1},\\
&\boldsymbol{\sigma} ^{\rm fr}(x)( \alpha _x, \beta _x)= J_ \varphi ^{-1}\mathbf{P} ^{\rm fr}( \varphi ^{-1}  (x) )( \alpha _x, T ^\ast _X \varphi ( \beta _x)),
\end{aligned}
\end{equation} 
for all $x \in \mathcal{D} $ and for all $ \alpha _x , \beta _x \in T^* _x \mathcal{D} $.
\medskip

From the expression of the free energy Lagrangian \eqref{FE_Lagrangian_NSF}, we deduce its spatial representation as
\begin{equation}\label{Lagrangian_NSF_Euler}
\ell( \mathbf{v} , \rho , T)=\int_ \mathcal{D}  \Lambda( \mathbf{v}, \rho , T) \mu _g= \int_ \mathcal{D} \frac{1}{2} \rho | \mathbf{v}|_g ^2\mu _g-\int_ \mathcal{D} \psi (\rho , T) \mu _g.
\end{equation}
Proceeding similarly as in \cite{GBYo2016b} we use the relations \eqref{def_Eulerian_variables} and \eqref{def_Eulerian_flux} to rewrite the variational formulation \eqref{VC_NSF_material}-\eqref{CV_NSF_material}-\eqref{KC_NSF_material} in spatial representation for a curve $( \mathbf{v} (t), \rho (t), \gamma (t), \sigma (t))$ as follows: 
\begin{equation}\label{VC_NSF_Eulerian} 
\delta \int_{t_1}^{t_2} \!\!\int_ \mathcal{D} \left[ \Lambda \big(\mathbf{v} , \rho , D_t\gamma \big)- \sigma D_t \gamma \right]  \mu _g dt=0, \quad\;\;\; \textsc{Variational Condition}
\end{equation} 
with respect to variations
\begin{equation}\label{variations} 
\delta \mathbf{v} = \partial _t \boldsymbol{\zeta} +[ \mathbf{v} , \boldsymbol{\zeta} ], \quad \delta \rho =- \operatorname{div}( \rho \boldsymbol{\zeta} ), \quad   \delta \gamma  , \quad \text{and} \quad   \delta \sigma,
\end{equation} 
subject to the constraint
\begin{equation}\label{CV_NSF_Eulerian} 
D_t\gamma  \bar D_ \delta   \sigma = ( \boldsymbol{\sigma} ^{\rm fr})^\flat : \nabla \boldsymbol{ \zeta }  -\mathbf{j} _S \cdot \mathbf{d} D_ \delta \gamma, \qquad\, \textsc{Variational Constraint}
\end{equation} 
with $ \boldsymbol{\zeta}  (t_i)=0$ and $ \delta \gamma (t_i)=0$, for $i=1,2$, where the curve $( \mathbf{v} (t), \rho (t), \gamma (t), \sigma (t))$ satisfies the constraint
\begin{equation}\label{KC_NSF_Eulerian}
D_t\gamma  \bar D_t\sigma = ( \boldsymbol{\sigma}  ^{\rm fr}) ^\flat : \nabla \mathbf{v}   -\mathbf{j} _S \cdot \mathbf{d} D_t \gamma+ \rho r, \;\;\, \textsc{Phenomenological Constraint}.
\end{equation} 

The first two expressions in \eqref{variations} are obtained by taking the variations with respect $ \varphi$, $ \mathbf{v} $, and $ \rho $, of the first two relations in \eqref{def_Eulerian_variables} and by defining the vector field $ \boldsymbol{\zeta} := \delta \varphi \circ\varphi^{-1} $. These formulas can be directly justified by employing the Euler-Poincar\'e reduction theory on Lie groups, for instance, see \cite{HoMaRa1998}.

In \eqref{VC_NSF_Eulerian}, \eqref{CV_NSF_Eulerian}, and \eqref{KC_NSF_Eulerian}, we introduced the notations
\[
\begin{array}{lll}
\vspace{0.2cm}&D_tf:= \partial _t f+ \mathbf{v}\cdot \mathbf{d}  f, & \qquad \bar D_t f := \partial _t f + \operatorname{div}( f  \mathbf{v} ),\\
\vspace{0.2cm}&D_ \delta f:= \delta  f+ \boldsymbol{\zeta} \cdot \mathbf{d}  f, &\qquad \bar D_ \delta  f := \delta  f + \operatorname{div}( f  \boldsymbol{\zeta}  ),
\end{array}
\]
for the Lagrangian time derivatives and variations of scalar fields and density fields.

\medskip

By applying \eqref{VC_NSF_Eulerian}, using the expression for the variations $ \delta \mathbf{v} $ and $ \delta \rho $, and $ \mathbf{v} |_{ \partial \mathcal{D} }=0$, $ \delta \gamma (t _i )=0$, $i=1,2$, we find the condition
\begin{align*} 
&\int_{t_1}^{t_2} \!\!\int_ \mathcal{D} \left[  \left( \frac{\partial  \Lambda  }{\partial \mathbf{v} }+ \left( \frac{\delta  \Lambda}{\delta T} - \sigma \right) \mathbf{d}  \gamma\right) \cdot (\partial _t \boldsymbol{\zeta} + [\mathbf{v} , \boldsymbol{\zeta}] )-\frac{\partial \Lambda}{\partial \rho } \operatorname{div}( \rho \boldsymbol{\zeta} ) \right.\\
&\left. \qquad\qquad\qquad \qquad\qquad \qquad \qquad \qquad  - \bar D_t\left( \frac{\delta \Lambda}{\delta T} - \sigma \right) \delta \gamma -\delta \sigma D_t\gamma \right] \mu _g\, dt=0.
\end{align*} 
Using the variational constraint \eqref{CV_NSF_Eulerian}, collecting the terms proportional to $ \boldsymbol{\zeta} $ and $ \delta \gamma $, and using $\boldsymbol{\zeta} |_{ \partial \mathcal{D} }=0$, $ \mathbf{v} |_{ \partial \mathcal{D} }=0$, $ \boldsymbol{\zeta} (t _i )=0$, $i=1,2$, we obtain the three conditions
\begin{align*} 
\boldsymbol{\zeta} : \quad &( \partial _t + \pounds _ \mathbf{v} ) \left( \frac{\partial \Lambda}{\partial \mathbf{v} }+ \left( \frac{\delta \Lambda}{\delta T} - \sigma \right) \mathbf{d}  \gamma\right)= \rho \,\mathbf{d}  \frac{\partial \Lambda}{\partial \rho }- \sigma \mathbf{d}  (D_t \gamma )+\operatorname{div} \boldsymbol{\sigma}  ^{\rm fr}  - (\operatorname{div}  \mathbf{j} _s )\,\mathbf{d} \gamma, \\
\delta \gamma : \quad & \bar D_t\left( \frac{\delta \Lambda}{\delta T} - \sigma \right) = - \operatorname{div}  \mathbf{j} _s \qquad\text{and}\qquad  \mathbf{j} _s\cdot \mathbf{n} ^\flat =0 \;\; \text{on $ \partial\mathcal{D} $} ,
\end{align*} 
where we introduced the Lie derivative notation $\pounds _ \mathbf{v} \mathbf{m}: = \mathbf{v} \cdot \nabla \mathbf{m} + \nabla  \mathbf{v} ^{\mathsf{T}} \cdot \mathbf{m} + \mathbf{m}\operatorname{div} \mathbf{v}  $ for a one-form density $ \mathbf{m}$ along a vector field $ \mathbf{v} $. Further computations and the phenomenological constraint \eqref{KC_NSF_Eulerian} finally yield the system
\begin{equation}\label{system_Eulerian} 
\left\{
\begin{array}{l}
\vspace{0.2cm}\displaystyle
( \partial _t + \pounds _ \mathbf{v} ) \frac{\partial \Lambda}{\partial \mathbf{v} }= \rho \,\mathbf{d}  \frac{\partial \Lambda }{\partial \rho }-   \frac{\partial \Lambda}{\partial T } \mathbf{d} T+ \operatorname{div} \boldsymbol{\sigma}  ^{\rm fr}\\
\vspace{0.2cm}\displaystyle  T \left( \bar D_t \frac{\delta \Lambda}{\delta T}  + \operatorname{div} \mathbf{j} _s \right)  = (\boldsymbol{\sigma} ^{\rm fr} ) ^\flat : \nabla \mathbf{v} - \mathbf{j} _s \cdot  \mathbf{d} T+ \rho r \\
\bar D_t \rho =0,
\end{array}
\right.
\end{equation} 
whose last equation, the mass conservation equation, follows from the definition of $ \rho $ in terms of $\varrho _{\rm ref} $.
These are the general equations of motion, in free energy Lagrangian form, for fluid dynamics subject to the irreversible processes of viscosity and heat conduction. By specifying this system to the Lagrangian \eqref{Lagrangian_NSF_Euler}, one immediately gets the Navier-Stokes-Fourier system in the form
\begin{equation}\label{NSF_Eulerian} 
\left\{
\begin{array}{l}
\vspace{0.2cm}\displaystyle
\rho ( \partial_t \mathbf{v} +\mathbf{v} \cdot \nabla \mathbf{v} )= - \operatorname{grad}  p +\operatorname{div} \boldsymbol{\sigma} ^{\rm fr}\\
\vspace{0.2cm}\displaystyle T\bar D_t \frac{\partial \psi }{\partial T} = \operatorname{div} (T\mathbf{j} _s ) -(\boldsymbol{\sigma} ^{\rm fr} ) ^\flat: \nabla \mathbf{v} - \rho r\\
\bar D_t \rho =0,
\end{array}
\right.
\end{equation} 
where $p= \rho  \frac{\partial \psi}{\partial \rho }- \psi $. The heat equation can be rewritten as
\[
T \left( \frac{\partial s }{\partial T  }D_tT  + \frac{\partial p}{\partial T} \operatorname{div} \mathbf{v} \right)  = (\boldsymbol{\sigma} ^{\rm fr} ) ^\flat: \nabla \mathbf{v} - \operatorname{div}( T \mathbf{j} _s ) +  \rho r ,
\]
where the partial derivatives are taken at constant mass density and constant temperature. In terms of usual coefficients$^{6}$ (the specific heat at constant volume $C_v$, the speed of sound $c_s ^2 $ and the diabatic temperature gradient $ \Gamma $, which may all depend on $ \rho $ and $T$) it reads
\[
\rho C_v \left(  D_tT+ \rho c_s ^2 \Gamma  \operatorname{div} \mathbf{v}\right) = (\boldsymbol{\sigma} ^{\rm fr} ) ^\flat : \nabla \mathbf{v} - \operatorname{div}( T \mathbf{j} _s ) +  \rho r.
\]
This heat equation is valid for any state equations. In the case of the perfect gas, it simplifies since $C_v$ is a constant and $ \rho ^2 c_s ^2  C_v\Gamma = p$. These results are summarized as follows.

\addtocounter{footnote}{1}
\footnotetext{We recall the expressions of the coefficients: $C_v=T\frac{\partial \eta }{\partial T}(\rho ,T)$, $ c_s ^2 =\frac{\partial p}{\partial\rho }(\rho , \eta )$, $\Gamma = \frac{\partial T}{\partial p}(p, \eta )$, where $\eta =s/ \rho $ is the specific entropy.}

\begin{theorem}[Free energy variational formulation for the Navier-Stokes-Fourier system -- spatial representation]
The Navier-Stokes-Fourier equations in spatial representation, given by
\begin{equation}\label{summary_NSF_Eulerian} 
\left\{
\begin{array}{l}
\vspace{0.2cm}\displaystyle\rho ( \partial_t \mathbf{v} +\mathbf{v} \cdot \nabla \mathbf{v} )= -\mathbf{d} p +\operatorname{div} \boldsymbol{\sigma} ^{\rm fr},\\[2mm]
\displaystyle\rho C_v \left(  D_tT+ \rho c_s ^2 \Gamma  \operatorname{div} \mathbf{v}\right) =  \boldsymbol{\sigma} ^{\rm fr} : \nabla \mathbf{v} - \operatorname{div}( T \mathbf{j} _s ) +  \rho r,
\end{array}
\right.
\end{equation}
with boundary conditions $\mathbf{v} |_{ \partial \mathcal{D} }=0$ and $\mathbf{j} _s \cdot \mathbf{n} ^\flat |_{ \partial \mathcal{D} }= 0 $, follow from the variational condition \eqref{VC_NSF_Eulerian} with the variational and phenomenological constraints \eqref{CV_NSF_Eulerian}, \eqref{KC_NSF_Eulerian}.
\end{theorem}

\begin{remark}[Thermodynamic phenomenology]{\rm
In order to close the system \eqref{summary_NSF_Eulerian}, it is necessary to provide phenomenological expressions  of the thermodynamic fluxes in terms of the thermodynamic affinities, compatible with the second law of thermodynamics. In our case, the thermodynamic fluxes are $ \boldsymbol{\sigma} ^{\rm fr}$ and $ \mathbf{j} _s$ and we have the well-known relations
\begin{equation*}\label{friction_stress_NSF} 
\boldsymbol{\sigma} ^{\rm fr}=2 \mu  (\operatorname{Def} \mathbf{v})^{\sharp }+ \left( \zeta - \frac{2}{3}\mu \right)(\operatorname{div} \mathbf{v} )  g^\sharp\quad\text{and}\quad T\mathbf{j} _s^{\flat}= - \kappa \mathbf{d}  T \;\; \text{(Fourier law)},
\end{equation*} 
where $ \operatorname{Def} \mathbf{v} = \frac{1}{2} (\nabla \mathbf{v} + \nabla \mathbf{v} ^\mathsf{T})$, $ \mu \geq 0 $ is the first coefficient of viscosity (shear viscosity), $ \zeta \geq 0$ is the second coefficient of viscosity (bulk viscosity), and $ \kappa \geq 0$ is the thermal conductivity. Generally, these coefficients depend on $ \rho $ and $T$.
}\end{remark}

\noindent \textbf{Constraints in infinite dimensions.} The variational formulation for the Navier-Stokes-Fourier system developed in this paper is based on \textit{nonlinear} and \textit{infinite dimensional} generalizations of the Lagrange-d'Alembert principle of nonholonomic mechanics. In the present case, the infinite dimensional constraint is \textit{not} associated to a mechanical constraint, it is the expression of the entropy production of the system.
For infinite dimensional constrained \textit{mechanical} systems, variational formulations have been used for example in 
 \cite{GBPu2012}, \cite{GBPu2015a} and \cite{GBPu2014}, \cite{GBPu2015b} to derive and study geometrically exact models for elastic strands with rolling contact and for flexible fluid-conducting tubes. An infinite dimensional generalization of the Lagrange-d'Alembert principle was proposed in \cite{GBYo2016a} to treat the case of 2$^{nd}$ order Rivlin-Ericksen fluids in the context of nonholonomic systems. We refer to \cite{ShBKZeBl2015} for a treatment of infinite dimensional constrained mechanical systems via Hamel's formalism.

\medskip

\noindent \textbf{Conclusion and future direction.} In this paper, we presented a Lagrangian variational formulation for the Navier-Stokes-Fourier system based on the free energy. This formulation is developed in a systematic way from the free energy variational formulation of the thermodynamics of discrete systems described in \S\ref{2_3}. The approach presented in this paper complements that made in \cite{GBYo2016b} as it uses the temperature, rather than the entropy, as an independent variable. The proposed free energy variational formulation is also well-adapted to include additional irreversible processes such as diffusion and chemical reactions treated in \cite{GBYo2016b}. It can also be extended to cover the case of moist atmospheric thermodynamics following \cite{FGB2017}.

\medskip

\noindent \textbf{Acknowledgements.} F.G.B. is partially supported by the ANR project GEOMFLUID, ANR-14-CE23-0002-01; H.Y. is partially supported by JSPS Grant-in-Aid for Scientific Research (26400408, 16KT0024, 24224004) and the MEXT ``Top Global University Project''.


\begin{thebibliography}{xx}
\bibitem[Appell(1911)]{Ap1911}
Appell, P [1911], Sur les liaisons exprim\'ees par des relations non lin\'eaires entre les vitesses, \textit{C.R. Acad. Sci. Paris}, \textbf{152}, 1197--1199.

\bibitem[Bloch(2003)]{Bl2003}
Bloch, A.~M. [2003], {\em Nonholonomic Mechanics and Control}, volume~24 of {\em Interdisciplinary Applied Mathematics}, Springer-Verlag, New York. With the collaboration of J. Baillieul, P. Crouch and J. Marsden, and with scientific input from P. S. Krishnaprasad, R. M. Murray and D. Zenkov.


\bibitem[Cendra, Ibort, de Le\'on, and Mart\'in de Diego(2004)]{CeIbdLdD2004}
Cendra, H., A. Ibort, M. de Le\'on, and D. Mart\'in de Diego [2004], A generalization of Chetaev's principle for a class of higher order nonholonomic constraints, \textit{J. Math. Phys.} \textbf{45}, 2785.

\bibitem[Chetaev(1934)]{Ch1934}
Chetaev, N.~G. [1934], On Gauss principle, \textit{Izv. Fiz-Mat. Obsc. Kazan Univ.},
\textbf{7}, 68--71


\bibitem[Ebin and Marsden(1970)]{EbMa1970}
Ebin, D.G. and J.E. Marsden [1970], Groups of diffeomorphisms and the motion of
an incompressible fluid, \textit{Ann. Math.} \textbf{92}, 102--163.



\bibitem[Gay-Balmaz(2017)]{FGB2017}
Gay-Balmaz, F. [2017], A variational derivation of the thermodynamics of a moist atmosphere with
irreversible processes, \url{https://arxiv.org/pdf/1701.03921.pdf}


\bibitem[Gay-Balmaz, Marsden, and Ratiu(2012)]{GBMaRa2012}
Gay-Balmaz, F., J.~E. Marsden, and T.~S. Ratiu [2012], Reduced variational formulations in free boundary continuum mechanics. {\it J. Nonlinear Sc.} \textbf{22}, 553--597.


\bibitem[Gay-Balmaz and Putkaradze(2012)]{GBPu2012}
Gay-Balmaz, F. and V. Putkaradze [2012], Dynamics of Elastic Rods in Perfect Friction Contact, \textit{Phys. Rev. Lett.} \textbf{109}, 244--303.

\bibitem[Gay-Balmaz and Putkaradze(2014)]{GBPu2014}
Gay-Balmaz, F. and V. Putkaradze [2014], Exact geometric theory for flexible, fluid-conducting tubes, \textit{C. R. M\'ecanique}, \textbf{342}, 79--84.
 
 
\bibitem[Gay-Balmaz and Putkaradze(2015a)]{GBPu2015a}
Gay-Balmaz, F. and V. Putkaradze [2015a], Dynamics of Elastic Strands with Rolling Contact, \textit{Physica D} \textbf{294}, 6--23.

\bibitem[Gay-Balmaz and Putkaradze(2015b)]{GBPu2015b}
Gay-Balmaz, F. and V. Putkaradze [2015b], On flexible tubes conducting fluid: geometric nonlinear theory, stability and dynamics, \textit{J. Nonlin. Sci.}, \textbf{25}(4), 889--936.

\bibitem[Gay-Balmaz and Yoshimura(2015)]{GBYo2015}
Gay-Balmaz, F. and H. Yoshimura [2015], Dirac reduction for nonholonomic mechanical systems on semidirect products, \textit{Adv. Appl. Math.}, \textbf{63}, 131--213.


\bibitem[Gay-Balmaz and Yoshimura(2017a)]{GBYo2016a}
Gay-Balmaz, F. and H. Yoshimura [2017a], A Lagrangian variational formulation for nonequilibrium thermodynamics. Part I: discrete systems, \textit{J. Geom. Phys.} \textbf{111}, 169--193.

\bibitem[Gay-Balmaz and Yoshimura(2017b)]{GBYo2016b}
Gay-Balmaz, F. and H. Yoshimura [2017b], A Lagrangian variational formulation for nonequilibrium thermodynamics. Part II: continuum systems, \textit{J. Geom. Phys.} \textbf{111}, 194--212.


\bibitem[Gay-Balmaz and Yoshimura(2017c)]{GBYo2017a}
Gay-Balmaz, F. and H. Yoshimura [2017c], Variational discretization for the nonequilibrium thermodynamics of simple systems, \url{https://arxiv.org/pdf/1702.02594.pdf}.


\bibitem[Gay-Balmaz and Yoshimura(2017d)]{GBYo2017c}
Gay-Balmaz, F. and H. Yoshimura [2017d], Dirac structures in nonequilibrium thermodynamics, \url{https://arxiv.org/pdf/1704.03935.pdf}



\bibitem[Holm, Marsden and Ratiu(1998)]{HoMaRa1998}
Holm, D. D., J. E. Marsden and T. S. Ratiu [1998], The Euler-Poincar\'e equations and semidirect products with applications to continuum theories, {\it Adv. in Math.} \textbf{137}, 1--81.


\bibitem[Marsden and Hughes(1983)]{MaHu1983}
Marsden, J.~E. and T.~J.~R. Hughes [1983], \textit{Mathematical Foundations of Elasticity} (Prentice Hall, New York, 1983) (reprinted by
Dover, New York, 1994).


\bibitem[Green and Naghdi(1991)]{GrNa1991}
Green, A. E. and P. M. Naghdi [1991], A re-examination of the basic postulates of thermomechanics, {\it Proc. R. Soc. London.}  Series A: {\it Mathematical, Physical and Engineering Sciences}, {\bf 432}(1885), 171--194.

\bibitem[Gruber(1997)]{Gr1997}
Gruber, C. [1997], \textit{Thermodynamique et M\'ecanique Statistique}, Institut de physique th\'{e}orique, EPFL. 

\bibitem[Gruber(1999)]{Gr1999}
Gruber, C. [1999], Thermodynamics of systems with internal adiabatic constraints: time evolution of the adiabatic piston, \textit{Eur. J. Phys.} \textbf{20}, 259--266. 

\bibitem[Marle(1998)]{Ma1998}
Marle, C.-M. [1998], Various approaches to conservative and nonconservative non-holonomic systems, \textit{Rep. Math. Phys.} \textbf{42}, 1/2, 211--229.

\bibitem[Pironneau(1983)]{Pi1983}
Pironneau, Y. [1983], Sur les liaisons non holonomes non lin\'eaires,
d\'eplacements virtuels \`a travail nul, conditions de Chetaev, Proceedings
of the IUTAM–IS1MMM Symposium on “Modern Developments in Analytical
Mechanics”, Torino 1982, \textit{Atti della Acad. della sc. di Torino}, \textbf{117}, 671--686.


\bibitem[Podio-Guidugli(2009)]{Po2009}
Podio-Guidugli, P. [2009], A virtual power format for thermomechanics,  
\textit{Continuum Mechanics and Thermodynamics}, \textbf{20}(8), 479--487.


\bibitem[Shi, Berchenko-Kogan, Zenkov, and Bloch(2015)]{ShBKZeBl2015}
Shi, D., Y. Berchenko-Kogan, D.~V. Zenkov, and A.~M. Bloch [2015], Hamel's Formalism for infinite-dimensional mechanical systems, \textit{J. Nonlin. Sci.}, \textbf{27}(1), 241--283.


\bibitem[Simo, Marsden, and Krishnaprasad(1988)]{SiMaKr1988}
Simo, J.~C., J.~E. Marsden and P.~S. Krishnaprasad [1988], The Hamiltonian structure of nonlinear elasticity: The material, spatial and convective representations of solids, rods and plates, \textit{Arch. Rational Mech. Anal.}, \textbf{104}, 125--183.



\bibitem[Stueckelberg and Scheurer(1974)]{StSc1974}
Stueckelberg, E.~C.~G. and P.~B. Scheurer [1974], \textit{Thermocin\'etique ph\'enom\'enologique galil\'eenne}, Birkh\"auser, 1974.


\bibitem[von Helmholtz(1884)]{He1884}
von Helmholtz, H. [1884], Studien zur Statik monocyklischer Systeme.
\textit{Sitzungsberichte der K\"{o}niglich Preussischen Akademie der Wissenschaften
zu Berlin}, 159--177.

\end{thebibliography}
\end{document}